# Global AI Ethics: A Review of the Social Impacts and Ethical Implications of Artificial Intelligence

Alexa Hagerty, PhD
Dovetail Labs; École des hautes études en sciences sociales, Paris. Master, Pratique de l'interdisciplinarité dans les sciences sociales, Chargé d'enseignement

Igor Rubinov, PhD
Dovetail Labs; Princeton University

**Abstract**

The ethical implications and social impacts of artificial intelligence have become topics of compelling interest to industry, researchers in academia, and the public. However, current analyses of AI in a global context are biased toward perspectives held in the U.S., and limited by a lack of research, especially outside the U.S. and Western Europe.

This article summarizes the key findings of a literature review of recent social science scholarship on the social impacts of AI and related technologies in five global regions. Our team of social science researchers reviewed more than 800 academic journal articles and monographs in over a dozen languages.

Our review of the literature suggests that AI is likely to have markedly different social impacts depending on geographical setting. Likewise, perceptions and understandings of AI are likely to be profoundly shaped by local cultural and social context.

Recent research in U.S. settings demonstrates that AI-driven technologies have a pattern of entrenching social divides and exacerbating social inequality, particularly among historically-marginalized groups. Our literature review indicates that this pattern exists on a global scale, and suggests that low- and middle-income countries may be more vulnerable to the negative social impacts of AI and less likely to benefit from the attendant gains.

We call for rigorous ethnographic research to better understand the social impacts of AI around the world. Global, on-the-ground research is particularly critical to identify AI systems that may amplify social inequality in order to mitigate potential harms. Deeper understanding of the social impacts of AI in diverse social settings is a necessary precursor to the development, implementation, and monitoring of responsible and beneficial AI technologies, and forms the basis for meaningful regulation of these technologies.

## INTRODUCTION

The ethical implications and social impacts of artificial intelligence are topics of compelling interest to industry, researchers in academia, and the public. However, current analyses of AI in a global context are biased toward perspectives held in the U.S., and limited by a lack of research, especially outside the U.S. and Western Europe. To effectively engage with the global community on issues related to the ethics of AI-driven technologies (indeed all digital technologies), it is essential to understand how these technologies are understood and implemented around the world.

The importance of a global conversation about the social impacts and ethics of AI has appeared in industry and government reports,[1] as well as in academic debates,[2] and is only beginning to receive sustained attention. This article offers an important step toward a meaningful conversation about AI ethics across cultures. Regional differences constitute a critical blind spot that has gone largely unexplored and represents an opportunity to advance the industry discussion on AI ethics. This path-breaking global research project offers a foundational framing of the debate and provides a review and synthesis of relevant social science research through perspectives and case studies from Africa, Asia, Latin America, the Middle East, as well as Southern and Eastern Europe.

A key contribution of this research asserts that in order to understand ethics, we must understand culture, and vice versa. Societies have unique ethical vocabularies, understandings, and expectations.[3] Terms like "fairness" and "privacy" can mean different things in different places.[4] This is not to say that all systems of values are created equal. Even the most cursory glance at human history confirms that societies have elaborated ethical systems that support human flourishing, but also systems that cause profound suffering. Ethics is not just a subject for study by philosophers or anthropologists. Nor is it simply principles that we encounter already existing in the world. Ethics is also something that we do. It requires action and engagement.[5] To understand AI ethics, we must think deeply about the relationship between society and technology across the globe. But deliberation is not enough. We must also commit to an active ethical engagement that ensures AI technologies support human flourishing around the world.

### KEY FINDINGS

**1) Regional differences are significant**
While scholarship on the global impacts of AI is lacking, social science research has a deep record of analyzing the relationship between society and technology more generally. This research suggests that AI is highly likely to have markedly different social impacts depending on cultural setting. Likewise, perceptions and understandings of AI are likely to be profoundly shaped by local cultural and social contexts.

**2) AI can exacerbate social inequality**
Research in U.S. settings demonstrates that AI-driven technologies have a pattern of entrenching social divides and exacerbating social inequality, particularly among historically-marginalized groups. Our research findings indicate that this pattern exists on a global scale, and suggest that

low- and middle-income countries may be more vulnerable to negative social impacts of AI and less likely to benefit from positive outcomes. Any amplification of social inequality can increase social instability, putting entire societies at risk, with potentially far-reaching geopolitical consequences.

**3) Further action is essential**

Despite the clear and evident need for global understandings of AI, research has been neglected. High quality, on-the-ground research is unavailable, resulting in major gaps in our understandings of the social impacts of these technologies around the world. Rigorous, independent ethnographic research is needed to investigate the ethical and social implications of AI across cultures. This is particularly critical to understand where AI systems may be amplifying social inequality and how harms may be mitigated.

**Methods**

This research project assessed the state of AI and related technologies in five global regions with particular focus on fourteen countries. A team of social science researchers reviewed more than 800 academic journal articles and monographs. Analysis was carried out in over a dozen languages and also assessed numerous policy papers, government reports, and local and regional media.[6]

Our analysis found major gaps in the scholarly literature. In fact, little systematic research has been done on the social impacts or ethical implications of AI systems anywhere in the world. Therefore, in order to think through these effects, we included consideration of adjacent technologies. As proxy technologies, these tools and systems are similar to or correspond to AI but are not explicitly aligned; examples include social media, cell phones, and forms of surveillance like CCTV. We believe that these technologies provide leading indicators of the social impacts of AI systems.

This article is a synthesis of a vast review of available literature in the social sciences. Assessment of academic databases suggests this is the first analysis of its kind: a large-scale, systematic review of the ethical implications and social impacts of artificial intelligence around the world.

## PART I: DEFINING THE TERMS

A serious conversation about AI ethics across cultures must begin with an interrogation of key terms: artificial intelligence, culture, and ethics. We argue that **a) artificial intelligence** is shaped by its social context at all phases of its development and use. As such, it takes distinct forms in different places. We propose a working definition of **b) culture** that avoids reductive thinking that can lead to stereotyping in favor of a dynamic and nuanced understanding of social context. We show how **c) ethics** are ultimately inseparable from culture. We demonstrate how ethnographic research methods are uniquely capable of analyzing ethics and culture in tandem.

### A) DEFINING AI: A Social View of AI

From an anthropological view, artificial intelligence is best understood as a "technosocial system," meaning that the technical aspects of AI are intrinsically and intimately connected to its social aspects. Social values and assumptions shape how we perceive, design, and use AI, as well as inform our perceptions, hopes, and fears of these technologies.[7]

**Perception and Imagination**

A subtle way that technology and society are woven together can be found in how we imagine technology. What hopes and fears drive the technologies we choose to develop, and how we accept, reject, and use the technologies around us?

Technologies emerge from a society's vision of the world and new technologies spark new ways to imagine the future.[8] When a society develops a technology, it does so because it has attained the technological mastery and know-how, but also because it has stories that inform the imaginations of makers. For example, Martin Cooper created the first personal cell phone after seeing Captain Kirk's communicator on an episode of *Star Trek*.[9]

A society's cultural imagination and its technologies are closely connected. For example, scholars have noted that the popularity of violent science fiction movies like the *Terminator* series has influenced how Americans imagine technology, leading to fears of killer robots.[10] By contrast, in Japanese cultural imagination, robots may be more likely to be associated with Mighty Atom (Astro Boy), a beloved manga character, perhaps leading to less hesitancy about coexistence with robots/AI.[11]

Some scholars, both Japanese and non-Japanese, attribute importance to Shintoism which recognizes all beings (including robots) as having spirits.[12] They theorize such cultural traditions lead to a preference for artificial intelligence technologies that take material form (hence the convergence of robotics and AI in Japan). Such an explanatory framework may account for why robots like Sony's AIBO and Honda's ASIMO were accepted so easily in Japan.[13]

However, many ethnographers and other scholars have recently noted that this ideal of the happy coexistence of robots and humans is not the whole story. For example, despite attempts to introduce robots as caregivers for the elderly to meet the challenge of Japan's aging population, most families are rather unwilling to utilize robots, preferring the human touch.[14] Thus we see that cultural imagination plays an important—but not always predictable—role in the relationship between technology and society. Ethnographic research can highlight the unexpected swerves and influences of cultural categories held in a collective imagination.

**Design and Development**

The social world in which AI is embedded shapes its development and design**.** People invent, design, and program technical systems. The general public has most often been alerted to the social aspects of AI design by dramatic problems and failures that make the news. But, even when AI systems function as intended, social values and cultural assumptions are always integrated into their design. When considering how development and design impact AI ethics across cultures, we must ask: **who makes AI and who is AI made for?**

1) **Who makes AI?** Who gets to participate in designing AI technologies? As will be discussed at greater length in Part II of the article , the persistent global digital divide excludes people from many parts of the world from participating in the design and development of AI technologies. For example, in many places, people lack the educational opportunities necessary to gain the specialized skills needed for AI.[15] Girls and women are especially affected by this skills gap.[16] These dispersed inequities raise the question: What are the long-term social consequences of AI technologies that are developed without full participation of women from the global south?

Even when people acquire skills and training to cross the digital divide, they may encounter obstacles. For example, African scholars persistently encounter visa issues that prevent them from participating in international conferences held in North America and Europe.[17] Visa issues led the International Conference on Learning Representation, a major AI symposium, to move their 2020 meeting to Addis Ababa, so that African researchers can participate.[18]

2) **Who is AI made for?** What cultural logics and assumptions are written into AI design? Data, the necessary foundation of AI, is deeply interwoven with society. People generate data. And people decide what counts as data and how, where, when, and why it should be gathered, sorted, and used. As Principal Researcher at Microsoft Research and founder of AI Now Kate Crawford explains, "data will always bear the marks of its history."[19]

From a cross-cultural perspective, a key concern is that much of the world leaves a thin digital footprint. People from low- and middle- income countries are likely to be radically underrepresented in the datasets central to developing AI systems, reinforcing the exclusion of their interests and needs. As a 2018 white paper from the World Economic Forum points out, an average U.S. household can generate a data point every six seconds. In Mozambique, where about 90% of people lack internet access, the average household generates zero digital data points.[20] In a world where data plays an increasingly powerful social role, to be absent from datasets may lead to increasing marginalization with far-reaching consequences.

**Implementation and Use**

The close relationship between the social and the technological informs how AI is designed, but more importantly how it is **utilized**. Technology is rarely used under laboratory conditions or by people with the same demographic profile as those who designed it or tested it.

Technologies enter a world that is already living, that is built on history, and that is shaped by economic and political structures.[21] As *MIT Technology Review*'s Karen Hao argues, "Even the fairest and most accurate systems can still be used to infringe on people's civil liberties."[22] The most thoughtfully-designed technologies can work in ways that are not just because they are used in the real world with all its imperfections and problems.

For example, in Saudi Arabia women cannot travel abroad without the signed consent of their male guardians (*muharram*)—husbands, fathers, brothers or adult sons. A 2015 ethnographic case study examined a mobile app intended to replace this infamous "yellow slip."[23] Initiated as part of the Saudi e-government strategy, the app alerted male relatives of women's movements through SMS messages. Journalist Safa Alahmad wrote, "The new compulsory text service, compliments of the Saudi ministry of interior, is not only a vicious reminder that Big Brother is watching me but that now he will snitch and tell my 'guardian' every time I leave the country."[24]

In this case, the technology itself was not necessarily biased or flawed, but it was deployed under existing social conditions with results that were both predictable (it enforced the status quo) and surprising (women found ways to speak up against the technology, in some cases by using other technologies, like Twitter).[25]

Technologies are used in the wild, in a complex and imperfect world. They can have unintended consequences and unexpected results. They can be used in ways that are creative and politically liberating or inequitable and repressive.

## B) DEFINING CULTURE

By now it should be clear that we cannot understand the impacts and ethics of AI without understanding the social world within which it is embedded. This requires grappling with the importance of culture. At first glance, the concept of "culture" may seem straightforward and intuitive. Yet when we begin to unpack its definition, culture turns out to be quite complex. The scholar Raymond Williams famously wrote that "culture is one of the two or three most complicated words in the English language."[26]

### Beyond Simple Slogans

Culture is often discussed in terms of ethnicity, nationality, and language. People from a particular culture are assumed to have a core, unchanging set of beliefs and characteristics. Harvard anthropologist and physician Arthur Kleinman names this prevalent form of talking about culture the "trait list approach."[27] Based on his immersive research in the United States and China, he critiques this common approach as far too simplistic. He warns that it leads to sweeping generalizations like the "Chinese believe this" and the "Americans believe that" — "as if entire societies or ethnic groups could be described by these simple slogans."[28] Anthropology rejects the "trait list approach" because it dangerously simplifies human experience, and leads to stereotyping rather than true understanding.

### The Problem with Culture

Anthropologists are wary of how oversimplified ideas about "culture" can be used in service of troubling political projects—like arguing for the innate superiority of one group of people over another and justifying racism and xenophobia.

For example, anthropologist Charles Briggs documented a cholera epidemic in an indigenous community in Venezuela.[29] Cholera is highly treatable, but the Venezuelan Ministry of Health failed to respond adequately and scores died. In response, officials blamed the deadly outbreak on "indigenous culture" which was portrayed in stereotypes and caricatures of dirty, poor, and ignorant *indígenas*. In this case, as in many others, "culture" was used to blame the victim, deflect responsibility from institutions, and place responsibility for a devastating public health crisis on people's supposed "way of life" rather than their access to clean water or adequate medical care.[30]

**Limits of the Culture Concept**
Even when “culture” is used with good intentions, problems can arise. Using a “trait list” approach can make it seem like “culture” is the answer to all questions of difference, distracting from other important considerations.

The limitations of “trait list” notions of culture have become very clear in medical settings. Physicians are increasingly committed to developing what they call “cultural competency,” or core skills of cross-cultural understanding and communication. However, studies reveal that physicians regularly make the mistake of attributing to “culture” issues that have other causes and require unrelated solutions.[31]

For example: a doctor in California is treating a recent Mexican immigrant who is HIV positive. The man's wife died of AIDS one year earlier, and their four-year-old son is also HIV positive. Despite the doctor’s attempts to explain the importance of regular medical treatment, the man has not been bringing his son to the clinic regularly for care. The doctor consults with a medical anthropologist to understand what cultural beliefs about health and medicine are preventing the man from seeking treatment for himself and his son. The anthropologist discovers that the man is a low-paid bus driver and he cannot afford to miss work to take his son to the clinic regularly. He misses appointments not because of cultural differences or distinct beliefs, but due to his practical socioeconomic situation.[32]

While such practical considerations may seem obvious, they often get overlooked in the name of culture. The insights gained from blind spots in “cultural competency” in medicine are important to issues of technology across cultures. Cultural differences exist alongside other differences (like economic status). This is true when we are thinking about technology, as well as medicine. Cultural differences will shape the way that people around the world respond to new technologies, but so will more everyday concerns, like literacy and broadband speed. To deeply understand people’s experiences we must move beyond oversimplified “trait list” approaches to culture.

**Culture: A Working Definition**
We propose a working definition of culture that honors the complex social lives of human beings but is also simple enough to be useful. Culture is a dynamic and ever-changing repertoire of shared understandings, inseparable from history, politics, and economics.

**1. Culture is shared**
Culture is a loosely-bounded system of shared experiences and understandings, some of which are explicit (consciously known, openly discussed, and present in public documents like rules and laws) and some of which are implicit (unconsciously known in the form of assumptions, intuition, and “common sense”). However, “shared” does not always mean agreed upon. Current debates in the U.S. about immigration and healthcare (sometimes called “the culture wars”) offer evidence that cultural experience can be shared without being agreed upon.

**2. Culture is heterogeneous**
Culture is not homogenous or seamless. It encompasses a diversity of subcultures, social groups, and individual differences. Cultural processes frequently differ even within close groups.

Differences in age, gender, political persuasion, class, education, religious conviction, ethnic identity, regional roots, and personality are just a few of the markers of diversity within a culture. For example, Vladimir Putin, Garry Kasparov, and Maria Sharapova are all Russian and share many cultural understandings, but their differences refute any easy notion of Russian culture as homogenous.

**3. Culture is always changing**
Culture is not static or permanent; it is always changing. Some aspects change quickly: music tastes, hairstyles, fashion trends, and diet fads. Other aspects change slowly: wealth disparity between black and white households in the U.S. has remained much the same since the 1950s.[33] The gender pay gap is entrenched.[34] Achieving diversity among corporate, political, and military elites has taken place at a glacial pace.[35]

**4. Culture is interconnected**
Culture is connected to and informed by political and economic structures, and inseparable from experiences of history. We can't understand German culture without understanding National Socialism and the Stasi. We can't understand South African culture without understanding colonialism and apartheid. We can't understand Irish culture without the potato famine and the Penal Laws banning Catholics from holding public office and buying land.

**5. Culture is reflexive**
Finally, we can only reflect on other cultures through the lens of our own culture and experience. Are apartments in Paris small? It depends on whether you are coming from Tokyo or Texas. Is Mexican pico de gallo salsa spicy? Not if you grew up in Tamil Nadu eating Chettinad cuisine. Is swimming topless at a public beach scandalous or expected? Do armed units patrolling the street give you a sense of safety or fear? Your answers depend on your cultural background and personal experience. There is no simple way to eradicate this cultural "baggage." The best cross-cultural thinkers and researchers embrace their own unique cultural background and use it as a tool. They don't try to think about culture in a vacuum, but understand it as a meeting of experiences.

**CASE STUDY: COLONIAL LEGACIES**

New technologies are tied to existing histories. For instance, the history of colonialism in Africa has created a continuing legacy of distrust. The late nineteenth century "Scramble for Africa" in which European powers fought to occupy territory and exploit resources on the continent continues to impact contemporary African-European collaborations.

In one case, an African research partnership for an anti-malarial drug disintegrated when local scientists affiliated with foreign drug companies received foreign patents.[36] Some traditional healers have become suspicious of scientists and potential intellectual property theft, and now refuse to share their herbal knowledge. Cases like this raise profoundly important questions of ownership, international patents, and the ongoing foreign appropriation of African resources and knowledge practices. What will this mean for AI developers and local concerns over intellectual property? How can foreign researchers navigate the boundaries between collaboration and exploitation?

**C) DEFINING ETHICS: A Plurality of Ethical Frameworks**
Western moral philosophy is built on three pillars: Aristotelian virtue ethics, Kantian deontology, and consequentialism. But formal moral frameworks have originated in societies across the world, including Confucian, Shinto, and Hindu thought, as well as the religious frameworks of Judaism, Christianity, and Islam, among many others.[37] Not only do a multitude of moral frameworks exist across cultures, there is significant variation within these frameworks, adding to their diversity. We need only consider the differences between Catholic and Protestant thought in Christianity, or Sunni and Shiite thought in Islam, to see that this is the case.

While formal moral frameworks drawn from philosophy and religion are important, they form only one aspect of a cross-cultural study of ethics. Every culture has a living system of ethics, which includes formal moral frameworks, but also many other factors such as politics, history, law, customs, common sense, and individual experience.[38] These multiple aspects of a culture's living system of ethics are linked together.[39] In other words, ethics and culture are inseparable.

To understand ethics, we must understand culture, and vice versa. Ethics and culture must be considered together as interlocking strands of social DNA. These twin helices constitute each other. To understand any culture, we must consider its values; to understand values, we must understand their cultural context. Because ethics and culture are joined, we cannot study ethics solely as a philosophical abstraction. Ethics must be studied in everyday cultural context to be fully understood. In other words, it is not enough to know the "rules of the game," we must also understand how people play.[40] Up-close and on-the-ground research is vital.

**Part I Conclusion**
In 'Part I: Defining the Terms,' we have defined the terms AI, culture, and ethics. We explained how **artificial intelligence** is shaped by social context at all phases of development and use. Our working definition recognizes **culture** as a vibrant and dynamic, not as a dry and dusty list of unchanging traits. We examined **ethics** as much more than just rules, but a living system. We argued that culture and ethics are ultimately inseparable.

## PART II: ASKING HARD QUESTIONS

Rather than offering a simple set of solutions, situating AI ethics within heterogeneous and malleable cultural milieus complicates the way AI technologies interface with global populations. To address questions of the social impacts and ethical implications of artificial intelligence technologies, technology companies and governments are increasingly developing ethical principles. In 2018, sixteen countries released national AI strategies, all of which included at least some stated ethical principles.[41] While such principles can offer an important first step in discussions about ethical AI across borders and cultures, they also raise difficult questions. How are terms like "fairness" and "privacy" understood in countries outside the United States? Is inclusive AI the same everywhere? Could a system empower people in one part of the world, and be disempowering somewhere else? How will we know how these technologies are playing out on the ground? Is understanding local legal statutes and relying on independent assessments from global NGOs enough?

There are no easy answers, but ethical deliberation requires that we not back away from hard questions. In this section, we will explore two hard questions:

1) How do AI principles translate linguistically and culturally across borders?
2) How do AI principles translate in practice around the world?

## AI PRINCIPLES IN TRANSLATION

*How do AI principles translate linguistically and culturally across borders?* When translating concepts like "fairness" and "privacy" across cultures, we can expect to encounter mistranslations and misunderstandings along the way. The value-laden terms used in AI principles do not seamlessly translate between languages. In 2017, a machine-translation error caused a Palestinian man's Facebook post of "good morning" to be rendered as "attack them," leading to his arrest by Israeli police.[42] Even as machine translation becomes ever-more sophisticated, and errors of this type less likely, language carries subtle meaning.

In a more nuanced case, Facebook's "like" button is translated as *curtir* in Brazilian Portuguese, which is closer to the English word for "enjoy." People's reluctance to "enjoy" negative events led to algorithmic filtering with political consequences for indigenous land rights activism in Brazil.[43] Anthropologists Rodrigo Ochigame and James Holston followed an indigenous collective in Mato Grosso do Sul struggling to regain control of their land from the powerful agribusinesses that control it. The group uses Facebook as its primary mode of outreach to the public, and often posts videos of violent acts carried out by private agribusiness militias and photos of the funerals of murdered indigenous activists. The researchers found that although many Facebook users found the posts to contain important information, they were reluctant to "enjoy" posts of violence and oppression. Because users did not "like" the posts, Facebook's filtering algorithm considered them unpopular and reduced their visibility. Consequently, land rights activists faced barriers in spreading their message through social media.

This was not a case of mistranslation, but an illustration of the subtle power of words and the nuanced meanings they carry. AI principles are inevitably value-laden terms, dense with significance. Such terms, even when thoughtfully translated, can have distinct connotations and meanings in different cultures.

### Culture and Understanding

"Privacy" is a term frequently mentioned in discussions of AI ethics. Yet, when ethical principles call for "privacy" what do people hear? The answer depends very much on who is listening. Conceptions of privacy differ by culture. An American visiting the Netherlands is likely to be struck by the lack of curtains on windows, with families eating dinner and watching TV in clear view of the street. Americans are likely to have been raised with the ideal that privacy is a natural right of the individual, whereas someone who grew up in China is more likely to have absorbed a notion of privacy as something that pertains to the family rather than the individual. In China, the concept of privacy appears to be shifting. While it was once seen in primarily negative terms, often closely related to ''selfishness," that has softened.[44]

In the U.S., conversations about General Data Protection Regulation (GDPR) revolve around "privacy," but in Europe, the discussion centers on "data protection." As Julie Brill, former

Commissioner of the Federal Trade Commission, and current Microsoft Corporate Vice President and Deputy General Counsel has pointed out, these are not merely semantic differences, but reveal distinct cultural conceptions and concerns about the relationship between individuals, governments, and corporations. When Americans talk about "privacy" instead of "data protection" in the EU, they often miss this subtle but important cultural difference.

Conceptions of privacy are shaped by culture and intersect with our understandings of the nature of personhood and the relationship between the individual and society.[45] The ethical ideas informing AI principles do not travel light, but come encumbered with social histories and cultural assumptions.

## AI PRINCIPLES IN PRACTICE

AI principles are not just words. They are intended to guide practice. In this section we ask: How do AI principles translate in practice around the world? Is an AI system "fair" or "inclusive" the same way everywhere?

Our literature review reveals a dearth of research on the social impacts of AI systems, especially in settings outside the U.S. and Western Europe. Without further rigorous ethnographic research, this remains an open question. However, existing social science research on the intersection of society and technology gives us important clues to understanding how AI principles may translate in practice.

### AI Everywhere

AI technologies can be expected to take shape in distinct ways in different places.[46] When big ideas and big technologies come into contact with local cultures, everything and everyone is changed by the encounter. Instead of thinking of big ideas and big technologies as being "brought" to a place, we should think of them as colliding with a place, and creating tensions, friction, and new possibilities. We can also think of technologies as being transplanted to a place, taking root in local soil.[47] The result may be an "invasive species" that causes unanticipated and long-lasting harm, or hardy hybrids that bear new fruit.[48]

We can be sure that as AI principles and systems travel, in every place they collide with a local culture, something new and different will happen. AI will take different forms in different regions.

We are seeing the fruits of AI technologies as they crop up around the world.

- In Brazil, AI is being used to decrease corruption by "Rosie," an AI system that detects and tweets unusual patterns of payment to politicians.[49]
- India is home to a quarter of tuberculosis cases across the world. In Delhi, a leading hospital is using AI to screen digital chest X-rays with remarkable success.[50]
- In Nigeria, machine learning is being used for "data-driven farming," to analyze soil and help farmers decide what crops to plant and how to manage them.[51]
- A start-up has developed an algorithm for use in classrooms in India, to gauge student comprehension of material by reading facial expressions.[52]

- In the São Paulo subway, a private company has installed a facial recognition system that detects people's emotional responses to ads.[53]
- In Rio, middle-class commuters check apps that monitor shootings before leaving the house.[54]
- The Delhi Police have started using predictive policing methods, analyzing satellite images and using clustering algorithms to locate "hotspots."[55]

What will be the social impacts of these AI projects? What fruit will they bear?

**CASE STUDY: BIOMETRIC IDENTITY PROGRAMS**

In recent years, several biometric identity systems have been rolled out in the developing world. ID4Africa, a self-described "movement," is a collaboration in progress between state governments, development agencies, and industry. The Aadhaar identity card in India is the world's largest program: as of 2018, 92% of India's resident population of 1.339 billion have received the card and been registered in this system.[56]

A principle argument in favor of biometric programs is that they support the universal procurement of formal identification, lack of which currently leaves large portions of the global poor unable to access government services or legally travel across borders. In addition, biometric programs are promoted as a means to centralize and expedite social services and enforce election integrity and the principle of one person, one vote.[57]

One of the most common arguments against biometric identification is the risk of surveillance and tracking through facial or voice recognition and the digital traces left by geolocation. Opponents argue that these could be used to intimidate and suppress political opposition.[58] Privacy experts voice concerns about storing sensitive data in centralized locations potentially vulnerable to hacking. Data breaches in the Aadhaar system attest to this risk.[59] In India, some people have dropped out of HIV antiretroviral treatment programs for fear that sensitive personal information will be exposed through Aadhaar, which is linked to a wide range of services.[60]

Critics have also pointed out that malfunctions of such vast systems can have dire repercussions. For example, the Aadhaar program has been linked to problems obtaining food rations, leading to several deaths by starvation.[61] Although biometric systems generally have a low error rate, at such a massive scale, even a 2% error rate could affect millions of people.[62]

**Amplifying Inequality**

While AI systems hold incredible promise for social goods like improved agriculture, better medicine, and more accessible education, so far this promise comes with a dark side. AI systems have a pattern of entrenching and amplifying social inequality.[63]

A few well-known examples from the U.S. illustrate this point: job recruiting tools biased against women;[64] Latinx and African American borrowers faced with discriminatory credit algorithms;[65] bias regarding race, gender, and/or sexual orientation in sentiment analysis systems, natural language processing technologies, and datasets of photos used to train image recognition software.[66] Taken together, these incidents reveal a pattern in which AI systems disproportionately affect historically disadvantaged, marginalized, and vulnerable groups.

### A Global Pattern

AI's pattern of entrenching and amplifying social inequality does not stop at national borders. In fact, countries outside the U.S. and Western Europe, particularly low- and middle-income countries, may be more vulnerable to the negative social impacts of AI systems, and less likely to benefit from positive outcomes. The World Economic Forum Global Future Council on Human Rights has determined that risks for discriminatory outcomes in machine learning are "especially high" for these countries.[67]

Nearly half of the world lives on less than $5.50 per day, political corruption is a significant problem for two-thirds of the global population, and more than a third of humanity lives under authoritarian rule.[68] AI systems are produced and used in imperfect, unequal landscapes in which any amplification of inequality could have profound and devastating effects.

We have already seen that stigmatized and persecuted groups, like religious, ethnic, sexual, and gender minorities, are particularly at risk. The role of social media distribution algorithms in the genocidal oppression of Myanmar's Rohingya population, [69] China's intensive surveillance of Uighur communities,[70] and the surveillance of political activists and dissidents by authoritarian-minded governments[71] are among recent well-known examples on the international stage. Any consideration of the ethics of AI across cultures must take into account how AI technologies may further entrench and amplify social inequalities.

**CASE STUDY: FACIAL RECOGNITION SYSTEMS AND POLICING IN BRAZIL**

"Fairness" is an important ideal in discussions of AI ethics. In the context of the United States, the use of AI in judicial sentencing and policing has received wide critical attention for bias and unfairness. As detailed in investigations from ProPublica and Georgetown Law Center, these technologies are least accurate for those they are most likely to affect: African Americans.[72]

Research on fairness and bias of AI systems in the U.S. context raises troubling questions for how such technologies will travel. As of now, there has been little rigorous, on-the-ground research on the social impacts of these technologies outside the U.S., so we can only speculate. But given what we know from the U.S. context, we should take the intersection of AI technologies and existing social biases in policing and sentencing very seriously.

A case in point: In January 2019, Brazilian President Jair Bolsonaro's ultraconservative Social Liberal Party (PSL) introduced a bill to permit security cameras with integrated facial recognition technology in public spaces for purposes of policing.[73] The same month, a government delegation visited China with an eye to buying facial recognition technology.[74]

### Deepening Digital Divides

Despite high global rates of connectivity, the digital divide persists.[75] The emergence of AI technologies may reinforce the current digital divide and introduce new forms of exclusion.[76]

Without adequate ICT infrastructure, communities around the world will be excluded from technological participation at a moment in history when AI could soon revolutionize all areas of daily life.[77] As a result, AI could exacerbate existing disparities and lead to further marginalization of communities. One of the drivers of this divide is exclusion from the datasets that AI systems are trained and built on. Much of the world leaves a thin digital footprint; people from low- and middle- income countries are likely to be radically underrepresented in the datasets central to developing AI systems, reinforcing this exclusion.

Furthermore, a lack of high-skill employment in large swaths of the world can leave communities out of the opportunities to redress errors or ethical missteps baked into the technological systems. Many people around the world lack educational opportunities to develop the specialized skills needed for AI; girls and women are especially affected.[78] This skills gap shuts out people from participating as creators (not just consumers or subjects) of AI technologies.[79]

**AI Labor Exclusions**

While AI is anticipated to fuel global economic growth, most of the gains are predicted in the U.S. and China.[80] Low- and middle-income countries are likely to reap only modest benefit, while also bearing the brunt of AI-driven job loss.[81] As a 2019 report from the ILO states succinctly: “Left to its current course, the digital economy is likely to widen both regional and gender divides.”[82]

Most AI tools and industries are likely to be concentrated in a handful of countries while the poorest countries will have very little chance of harnessing these technologies for their own domestic economies. According to PwC, of the $15.7 trillion in wealth AI will generate globally by 2030, seventy percent will accrue to China and America.[83]

Analyses predict that AI and other new technologies will continue to benefit higher-skilled workers who can utilize creativity, problem-solving, and interpersonal skills.[84] Low- and medium-skilled workers are expected to face downward pressures from increasingly competent machines and AI software. There is a strong possibility that these downward pressures will exacerbate already high levels of income inequality globally.[85] For example, low- and middle-income countries are seeing the rapid expansion of an AI-driven informal labor sector, leading to what critics call a rising class of “digital day laborers,”[86] such as poorly paid “clickworkers” who trace and label photos to support machine learning.[87] The Internet Society warns, “There is a high risk that the benefits of AI will be unevenly distributed within and across societies—exacerbating current and future digital divides” with geopolitical implications.[88]

**Global Risks**

The social impacts of AI systems will be distinct in different places. The repercussions of these impacts have the potential to be widely, even globally, felt. While the negative consequences of AI systems will be felt by the most vulnerable groups first, and most directly, any amplification of social inequality can increase social instability, and put entire societies at risk.

Scholarly literature across disciplines highlights troubling links between social inequality, political and economic instability, authoritarian forms of government, civil unrest, and violent conflict.[89] Social issues of this magnitude can have widely-felt repercussions, for example by increasing global migration or disrupting international financial markets.[90] Therefore, even if the most troubling negative effects of AI systems are initially felt by only a small portion of a population, this does not mean that ill-effects will be contained to one social group. Any increase in social inequality carries risks for entire societies.

**Social Organization and Control**

AI poses hard questions. At a meeting of the UN High-Level Panel on Digital Cooperation in January 2019, Assistant Secretary-General for Strategic Coordination Fabrizio Hochschild reflected on these risks and uncertainties:

> "Will emerging technologies contribute to peace overall or will they undermine it? Will they generally further access to sustainable development or will they further inequality? Will they facilitate respect for human rights or will they provide new tools to those who wish to contain or violate the realization of human rights?"[91]

As AI principles are translated and put into practice around the world, these are difficult questions to confront. AI systems offer powerful tools of social organization and control. AI will be leveraged in many pro-social ways: from guaranteeing the integrity of elections, to increasing the efficiency of public transportation systems, to more effectively responding to natural disasters. However, AI systems can also be used for non-democratic ends, such as intensive surveillance of citizens and to intimidate activists and dissidents.[92]

Social control is exercised not only by what governments *can* do but what a population *believes* they can do. Thus, governments routinely both exaggerate and minimize the sophistication of their surveillance capabilities in order to appear more powerful or less threatening, depending on their aims. For example, during the Obama administration, the U.S. government exaggerated the success of the National Security Agency's surveillance programs in averting terrorist attacks.[93] Likewise, the Chinese government routinely overstates its capabilities to appear more omnipotent, evidenced by a widely circulated story of the unlucky concertgoer caught for "economic crimes" by a facial recognition system among a crowd of 60,000 at a pop concert.[94]

The reality is often more modest. Surveillance is patchy, and much of the technology is not yet as powerful as it appears. As *The New York Times* reported, although police wear sleek smart glasses, the facial recognition technology they use requires someone to stand still for several seconds to make an identification; they are mostly used to check identity not nab crooks.[95] The infamous billboards displaying the faces of jaywalkers are the result of humans sifting through photos taken at crosswalks, not facial recognition.[96] Social control relies not only on technological sophistication, it is also closely connected to cultural imagination.

**CASE STUDY: SINGAPORE, SURVEILLANCE, AND SARS**

In Singapore, state surveillance is widely accepted as a fact of life.[97] For example, the telecommunications firm SingTel has been criticized since the early 2000s for its role in mass surveillance, but this appears to generate little controversy among Singaporeans, where state surveillance in the form of CCTV, monitoring of social media, and "lateral surveillance" by peers appear to be an established norm.[98]

Singapore's response to the severe acute respiratory syndrome (SARS) outbreak in 2003 revealed the extent of state authority and the political will for surveillance and social control. The Singapore government enforced exceptional measures, including: blanket screening of entering visitors, school closures, placing 8,000 people under home quarantine (including extensive video monitoring by a security company), placing 4,300 people under telephone surveillance, and issuing electronic wrist tags for quarantine violators.[99] While Singapore had the third largest rate of infection in Asia, it managed to contain the disease in just over three months. Not only did the population comply with the extreme measures, many observers see the government's handling of SARS as having increased national pride and allegiance to the state.[100]

**Tolerance, Acceptance, and Resistance**

As AI technologies become more pervasive and their social impacts more deeply felt, we can expect a range of responses. In any given society, facial recognition technologies are likely to be embraced by some people and greeted as a threat by others. Some people will find it convenient not to have to use a card at the ATM or feel safer in public spaces with police surveillance. Other people will worry about their privacy, immigration documentation, or parole status. The social impacts will not be "one size fits all" even within the same country or city.

Likewise, societies will evaluate the ethics and social impacts of AI systems differently. While some societies appear poised to embrace or at least tolerate wide-scale government surveillance enabled by AI systems, in other places in the world these technologies are sparking resistance.

In countries like Singapore and China, surveillance, whether AI-driven or in other forms, does not seem to generate much controversy among citizens. State surveillance appears to be an acceptable exchange for security and stability.[101] However, in other regions there have been signs of discontent. In 2016, Venezuela introduced a national ID system, "carnet de la patria," or "fatherland card."[102] Developed with support from Chinese telecom ZTE Corp, the card tracks users across a range of services and is linked to healthcare and social programs, including subsidized food. Shortly after its introduction, activists hacked into the national database and deleted accounts of prominent politicians in an act of political defiance.[103]

A 2019 study examining the ubiquitous presence of cameras and social media in Israel and the Occupied Palestinian Territories noted that while Israeli state surveillance is aimed at tracking Palestinian dissidents and alleged terrorists, the faces of Israeli soldiers are also captured and circulated on social media, giving rise to demands for personal accountability. This study demonstrates that surveillance intended as top-down may move in bottom-up directions, being used by people under surveillance in unexpected ways.[104]

In Part I of the article, we discussed a case in which Saudi Arabia replaced "yellow slips" requiring male consent for women's travel with an app. While this demonstrates how states can use technologies to monitor and constrict citizen movement, it also illustrates how ordinary women contested new forms of technologically-mediated surveillance, for example when activist

Manal Al-Sharrif alerted her Twitter followers whenever her husband received a text notification about her movements.[105]

In Singapore and China too, cracks in acceptance appear to be growing as individuals have greater access to outside media and economic growth begins to slow. As one observer writes, "The more time Singaporeans spend online, the more they read, the more they share their thoughts with each other and their government, the more they've come to realize that Singapore's light-touch repression is not entirely normal among developed, democratic countries—and that their government is not infallible."[106] These realizations are not only emerging among a more wary populace but among the firms and agencies producing the technologies. An article on the official WeChat channel of Tencent Digital in China on "positive and negative aspects of facial recognition technology" discusses the ethical dilemmas involved with the development of facial recognition technology by major companies outside of China, noting that "all technology is a double-edged sword." Though the December 2018 post could have been tailored to garner positive public relations, the article marks an important and unusual public consideration of AI ethics in China. A consideration reinforced by the introduction of the Beijing AI Principles in May of 2019, further underscoring a "surprising willingness to discuss such issues within Chinese policy circles."[107]

**CASE STUDY: SOCIAL CREDIT SYSTEM IN CHINA**

One of the most widely discussed developments of life in contemporary China is also one of the least clearly understood abroad. Almost every day, a news article is published alleging that China's social credit scoring system ranks citizens based on their political views and social behavior, drawing data from facial recognition-enabled surveillance cameras as well as e-commerce and social media platforms that seamlessly integrate public and private sector data. Yet many of these core assumptions about the social credit system do not match facts on the ground: there is no unified national "social credit score," facial recognition and AI are not included in evaluated data, and the information shared between the state and private sector is currently patchy and governed by limited memoranda of understanding.

The "social credit system" is a blanket term for a series of initiatives meant to strengthen the enforcement of pre-existing laws in China.[108] Auxiliary objectives include the creation of a reliable financial credit system, elimination of market fraud, and the reduction of "dishonest" behavior in fields as diverse as environmental pollution and academic plagiarism. Although numerical scores are used to evaluate the performance of individuals and companies in some limited and still experimental city-level cases, the state itself is not currently using the system to make socially consequential judgments about citizens.[109]

Overall, the Chinese public generally perceives the social credit system as beneficial to society. The system is well received by Chinese citizens, who believe it is an example of the state taking action to prevent fraud (although there are few accounts from blacklisted entities available by which to analyze this perspective). Surveys have found that younger, better-educated urban residents were most inclined to favor the system, which the authors suggest may result from this sub-population benefiting the most from good credit.[110]

China's social credit system is currently a patchwork of public-private partnerships piloted in various second-tier cities. It is not yet the sinister dragnet that Western media fear, nor is it a ruse. These tools offer a set of population monitoring and evaluation metrics that are still a work in progress, with potentially far-reaching consequences in store.

## CONCLUSION

In our review and synthesis of social science research in five world regions with particular focus on fourteen countries, our analysis finds that AI can be expected to have critical social impacts around the world, and that these impacts will have significant regional variation. Likewise, perceptions and understandings of AI are likely to be profoundly shaped by cultural context. Consequently, further research will be required to make sense of the social impacts and ethical implications of emerging and future AI technologies.

### The Case for Ethnography

AI systems are created and used in a flawed, imperfect, and unpredictable world. It is the meeting of AI technologies and the real world that demands our attention, and must be a research priority. We call for further research on the social and ethical impacts of AI, through ethnographic inquiry that is attuned to how these technologies are affecting communities in all parts of the world.

Kate Crawford of AI Now has long advocated for empirical research to assess the social impact of AI technologies at all stages, from design to deployment and regulation.[111] Data & Society, founded by Microsoft Research Principal Researcher danah boyd, has recently established an initiative to support "empirically grounded research rather than speculation" because the social impacts of AI "can only be fully understood by observing, listening, and speaking with people on the ground."[112]

As AI principles and AI practices travel around the world, we—thinkers and makers—must travel with them. We must closely engage with them, whether they are being used across the globe or across the street. We must track them at every step from design to daily use, paying close attention to how they are translated, understood, and implemented. We must not stop asking hard questions, or accept easy answers. If we do this, AI systems can work to increase human flourishing around the globe rather than amplify inequality.

This article has argued that AI systems have demonstrated a pattern of exacerbating inequality, often in the most unequal societies and particularly for the most vulnerable populations. While AI-based technologies offer enormous promise to better the lives of people everywhere, they also bring risks. The unwanted social impacts of AI may be felt most immediately by historically marginalized groups, but they are unlikely to be limited to any particular group. Their repercussions may be widely felt. The most socially vulnerable populations are the canaries in the data mining, so to speak. As Virginia Eubanks has so trenchantly observed, if we want to predict the future of technological impacts on society, we need only train our eyes on how these systems affect the most vulnerable among us: "Systems tested in low rights environments will eventually be used on everyone."[113]

**Toward an AI Ethics across Cultures**

We need an ethics of AI across cultures that takes these risks seriously. As philosopher and scholar of the Holocaust John K. Roth has written, talking about ethics is not enough: "no simple reaffirmation of ethics, as if nothing disastrous had happened, will do."[114] We must act in full knowledge that AI systems are developed and used in an imperfect and unequal world—in a world where disastrous things have happened, and can happen again.

Talking about AI ethics is not enough, and principles are only a first step. A practice of ethics is crucial. We must work to create the world we want. "What is our human project for the digital age?" asks Luciano Floridi, Professor of Philosophy and Ethics of Information at Oxford, echoing the increasingly urgent call to think not only about the technology we *can* produce, but about the technology we *should* produce.[115]

We see cause for deep concern regarding the pattern of amplifying inequality that AI systems have exhibited so far. However, AI could also be a force for decreasing divisions and creating a more equal world. "AI can help to bridge the digital divide and create an inclusive society," as Uyi Stewart of the Bill & Melinda Gates Foundation has said.[116] But to achieve this "puts an enormous responsibility on AI practitioners to be ethical, transparent, and intentional in how we implement AI technologies. We need to pay attention to practical challenges on the ground in executing on the promise of AI."[117]

To determine what technologies we should produce, to implement our technologies with intention, to monitor their impact on human lives everywhere on earth, and to evaluate whether they are supporting or undermining the societies we want to build for ourselves and future generations—this is the ethical project that lies before us.

**Funding:** This research was funded by Microsoft

Acknowledgements: Literature review research was undertaken by the following scholars:

Shazeda Ahmed, PhD Candidate, University of California, Berkeley, School of Information
Leah M. Ashe, PhD, Cardiff University, Planning and Geography; Postdoctoral scholar, University of Notre Dame's Center for Science, Technology, and Values
Bianca Vienni Baptista, PhD, Postdoctoral researcher, ETH Zürich (Switzerland), USYS Transdisciplinary Lab (TDLab)
Renata Barreto, JD/PhD Candidate, University of California, Berkeley, Jurisprudence & Social Policy
Marten Boekelo, PhD, University of Amsterdam, Amsterdam Institute for Social Science Research
Bidisha Chaudhuri, Assistant Professor, International Institute of Information Technology, Bangalore
Ivana Damnjanović, PhD, University of Belgrade, Political Theory
Jesse Dart, PhD Candidate, University of Sydney, Social Anthropology
Gabriele de Seta, PhD, Hong Kong Polytechnic University, Sociology
Zoe Hatten, PhD Candidate, Australian National University, Anthropology
Torin S. Jones, PhD Candidate, Stanford University, Anthropology
Eirini Malliaraki, Alan Turing Institute; MSc, Imperial College London; MA, Royal College of Art, Design, Engineering
Young Su Park, PhD, Stanford University, Anthropology; Postdoctoral researcher, Freie Universität Berlin
Kalpana Shankar, Professor, University College Dublin; School of Information & Communication Studies
Joanna Steinhardt, PhD, University of California, Santa Cruz, Anthropology
Tomo Sugimoto, PhD, Stanford University, Anthropology; Postdoctoral scholar, Yale University

[1] Walker, Summer. "We need a global conversation on AI in policing." *UN-TOC Watch: Global Initiative Against Transnational Organized Crime* https://globalinitiative.net/global-conversation-on-ai-in-policing/ accessed July 15, 2019; Bossman, Julia. "Top 9 ethical issues in artificial intelligence" *World Economic Forum* (October 21, 2016) https://www.weforum.org/agenda/2016/10/top-10-ethical-issues-in-artificial-intelligence/ accessed July 15, 2019; UNESCO Paris "Roundtable on 'Artificial Intelligence: Reflection on its complexity and impact on our society'" (September 11, 2018).

[2] Williams, Betsy Anne, Catherine F. Brooks, and Yotam Shmargad. "How algorithms discriminate based on data they lack: Challenges, solutions, and policy implications." *Journal of Information Policy* 8 (2018): 78-115; Wahl, Brian, Aline Cossy-Gantner, Stefan Germann, and Nina R. Schwalbe. "Artificial intelligence (AI) and global health: how can AI contribute to health in resource-poor settings?." BMJ global health 3, no. 4 (2018): e000798; Baum, Seth D. "Social choice ethics in artificial intelligence." AI & SOCIETY (2017): 1-12; Crawford, Kate, and Ryan Calo. "There is a blind spot in AI research." Nature News 538, no. 7625 (2016): 311; Mantelero, Alessandro. "AI and Big Data: A blueprint for a human rights, social and ethical impact assessment." Computer Law & Security Review 34, no. 4 (2018): 754-772.

[3] Strathern, Marilyn. The gender of the gift: problems with women and problems with society in Melanesia. No. 6. Univ of California Press, 1990; Linklater, Andrew. The transformation of political community: ethical foundations of the post-Westphalian era. Univ of South Carolina Press, 1998; Lebow, Richard Ned. The politics and ethics of identity: In search of ourselves. Cambridge University Press, 2012.

[4] Agre, Philip E., and Marc Rotenberg, eds. *Technology and privacy: The New landscape*. Mit Press, 1998; Bélanger, France, and Robert E. Crossler. "Privacy in the digital age: a review of information privacy research in information systems." MIS quarterly 35, no. 4 (2011): 1017-1042; Hartman, Laura P. "Technology and ethics: Privacy in the workplace." Business and Society Review 106, no. 1 (2001): 1-27.

[5] As philosopher John K. Roth, a scholar of the Holocaust writes, ethics is defined by "deliberation about the difference between right and wrong, encouragement not to be indifferent toward that difference, resistance against what is wrong, and action in support of what is right." Roth, John K. *The Failures of Ethics: Confronting the Holocaust, Genocide, and Other Mass Atrocities*. Oxford University Press, USA, 2015.

[6] Research took place over nine weeks between December 2018 and February 2019. Researchers produced initial regional and country reports of 30-60 pages in length, which were then analyzed and synthesized by a core team of researchers.

[7] Beer, David. "Power through the algorithm? Participatory web cultures and the technological unconscious." *New Media & Society* 11, no. 6 (2009): 985-1002; Zuboff, Shoshana. "Big other: surveillance capitalism and the prospects of an information civilization." *Journal of Information Technology* 30, no. 1 (2015): 75-89; Lash, Scott. "Power after hegemony: Cultural studies in mutation?." *Theory, culture & society* 24, no. 3 (2007): 55-78; Latour, Bruno. *Reassembling the social: An introduction to actor-network-theory*. Oxford university press, 2005; Katz, Yarden, Manufacturing an Artificial Intelligence Revolution (November 27, 2017). https://ssrn.com/abstract=3078224 or http://dx.doi.org/10.2139/ssrn.3078224

[8] Heidegger, Martin. *The Question Concerning Technology and Other Essays*. Harper and Row, New York, 1977. Ihde, Don. "Philosophy of technology." In *Philosophical problems today*, pp. 91-108. Springer, Dordrecht, 2004; Verbeek, Peter-Paul. *What things do: Philosophical reflections on technology, agency, and design*. Penn State Press, 2005.

[9] Bryant, Jennings, Susan Thompson, and Bruce W. Finklea. *Fundamentals of media effects*. Waveland Press, 2012.

[10] Richardson, Kathleen. *An anthropology of robots and AI: annihilation anxiety and machines*. Routledge, 2015.

[11] Robertson, Jennifer. *Robo Sapiens Japanicus: Robots, Gender, Family, and the Japanese Nation*. University of California Press, 2017.

[12] Geraci, Robert. 2006. "Spiritual Robots: Religion and Our Scientific View of the Natural World." *Theology and Science* 4(3): 229-246; Katsuno, Hirofumi. 2011. "The Robot's Heart: Tinkering with Humanity and Intimacy in Robot-Building." *Japanese Studies* 31(1): 93-109; Kitano, Naho. 2012. "'Rinri': An Incitement towards the Existence of Robots in Japanese Society." *International Review of Information Ethics* 6: 78-83; Allison, Anne. "The Japan fad in global youth culture and millennial capitalism." *Mechademia* 1, no. 1 (2006): 11-21; Jensen, Casper Bruun, and Anders Blok. "Techno-animism in Japan: Shinto cosmograms, actor-network theory, and the enabling powers of non-human agencies." *Theory, Culture & Society* 30, no. 2 (2013): 84-115.

[13] Kubo, Akinori. 2013. "Plastic Comparison: The Case of Engineering and Living with Pet-Type Robots in Japan" *East Asian Science, Technology and Society: An International Journal* 7: 205–220; Kubo, Akinori. 2015. *The*

*Anthropology of Robots: Automatons and Humans in Twentieth-Century Japan* (*Robotto no jinruigaku: nijusseiki nihon no kikai to ningen*). Kyoto: Sekai shisō sha.

[14] Wright, James. 2018. "Tactile Care, Mechanical Hugs: Japanese Caregivers and Robotic Lifting Devices." *Asian Anthropology* 17(1): 24-39.

[15] Bughin, Jacques, Jeongmin Seong, James Manyika, Michael Chui, and Raoul Joshi. "Notes from the AI frontier: Modeling the impact of AI on the world economy." Discussion Paper, McKinsey Global Institute, September 2018. https://www.mckinsey.com/featured-insights/artificial-intelligence/notes-from-the-AI-frontier-modeling-the-impact-of-ai-on-the-world-economy

[16] World Economic Forum, "Global Gender Gap Report 2018." http://reports.weforum.org/global-gender-gap-report-2018/assessing-gender-gaps-in-artificial-intelligence/?doing_wp_cron=1551634743.8178520202636718750000EY Global 2018.

[17] In November 2017, influential AI scholar Yoshua Bengio spoke to this perennial problem in an interview with MIT Technology Review's Will Knight: "In Europe or the U.S. or Canada it is very difficult for an African researcher to get a visa. It's a lottery, and very often they will use any excuse to refuse access. This is totally unfair. It is already hard for them to do research with little resources, but in addition if they can't have access to the community, I think that's really unfair." ***https://www.technologyreview.com/s/612434/one-of-the-fathers-of-ai-is-worried-about-its-future/***

[18] Johnson, Kari. "Major AI Conference Is Moving to Africa in 2020 Due to Visa Issues." Venture Beat. 2018. https://venturebeat.com/2018/11/19/major-ai-conference-is-moving-to-africa-in-2020-due-to-visa-issues/.

[19] Rosenberg, Scott. "Why AI is Still Waiting for its Ethics Transplant." Wired Magazine, November 1, 2017. https://www.wired.com/story/why-ai-is-still-waiting-for-its-ethics-transplant/

[20] There are geographical and gendered dimensions to data exclusion. Rural areas are more affected, as are women. See: World Economic Forum White Paper, "How to Prevent Discriminatory Outcomes in Machine Learning." March 12, 2018. https://www.weforum.org/whitepapers/how-to-prevent-discriminatory-outcomes-in-machine-learning; "Internet of Things, Privacy and Security in a Connected World," FTC, https://www.ftc.gov/system/files/documents/reports/federal-trade-commission-staff-report-november-2013-workshop-entitled-internet-things-privacy/150127iotrpt.pdf; "Bridging the Gender Gap: Mobile Access and Usage in Low- and Middle- Income Countries," GSMA, https://www.gsma.com/mobilefordevelopment/wp-content/uploads/2016/02/GSM0001_03232015_GSMAReport_NEWGRAYS-Web.pdf; "Rural Coverage: Strategies for Sustainability," GSMA, https://www.gsmaintelligence.com/research/?file=53525bcdac7cd801eccef740e001fd92&download

[21] Jasanoff, Sheila. "Virtual, visible, and actionable: Data assemblages and the sightlines of justice." *Big Data & Society* 4, no. 2 (2017); Latour, Bruno. *Reassembling the social: An introduction to actor-network-theory*. Oxford University Press, 2005.

[22] Hao, Karen. "Making Face Recognition Less Biased Doesn't Make It Less Scary." MIT Technology Review, January 29, 2019 https://www.technologyreview.com/s/612846/making-face-recognition-less-biased-doesnt-make-it-less-scary/.

[23] Johnson, Mark. 2015. "Surveillance, pastoral power and embodied infrastructures of care among migrant filipinos in the kingdom of Saudi Arabia". *Surveillance and Society.* 13 (2): 250-264.

[24] Harding, Luke. "Saudi Arabia criticised over text alerts tracking women's movements." The Guardian, November 23, 2012. https://www.theguardian.com/world/2012/nov/23/saudi-arabia-text-alerts-women.

[25] The relationship between technology and social activism is far more complicated than simple technological empowerment. As sociologist Zeynep Tufekci has articulated, it is a story of "entwined fragility and empowerment." *Twitter and tear gas: The power and fragility of networked protest.* Yale University Press, 2017.

[26] Williams, Raymond. *Keywords: A Vocabulary of Culture and Society*. Routledge, 2013.

[27] Kleinman, Arthur, and Peter Benson. 2006. "Anthropology in the clinic: the problem of cultural competency and how to fix it." PLoS Med 3(10): e294.

[28] Ibid.

[29] Briggs, Charles L. and Clara Mantini-Briggs. *Stories in the time of cholera: Racial profiling during a medical nightmare*. University of California Press, 2003.

[30] Ibid.; Farmer, Paul E., Bruce Nizeye, Sara Stulac, and Salmaan Keshavjee. "Structural violence and clinical medicine." *PLoS medicine* 3, no. 10 (2006): e449.

[31] Metzl, Jonathan M., and Helena Hansen. "Structural competency: Theorizing a new medical engagement with stigma and inequality." *Social Science & Medicine* 103 (2014): 126-133; Kleinman, Arthur, and Peter Benson. 2006. "Anthropology in the clinic: the problem of cultural competency and how to fix it." PLoS Med 3(10): e294.

[32] Ibid.

[33] In 1953, the net worth of the typical black household was 20% of that of the typical white household. In 2013, the wealth of the median black family in the US had fallen to 10% that of its white counterpart. Guilford, Gwynn. "Black income is half that of white households in the US—just like it was in the 1950s." Quartz. September 1, 2018. https://qz.com/1368251/black-income-is-half-that-of-white-households-just-like-it-was-in-the-1950s/; Kuhn, Moritz, Moritz Schularick, and Ulrike Steins. "Income and Wealth Inequality in America, 1949-2016." *Opportunity & Inclusive Growth Institute Working Paper* (2018). https://www.minneapolisfed.org/institute/institute-working-papers/income-and-wealth-inequality-in-america-1949-2016; National Bureau of Economic Research, "Household Wealth Trends in the United States, 1962 to 2016: Has Middle Class Wealth Recovered?" November, 2017. https://www.nber.org/papers/w24085.pdf.
[34] Graf, Nikki, Anna Brown, and Eileen Patten. "The narrowing, but persistent, gender gap in pay." *Pew Research Center,* April 9 (2018). https://www.pewresearch.org/fact-tank/2019/03/22/gender-pay-gap-facts/
[35] Zweigenhaft, Richard L., and G. William Domhoff. *Diversity in the Power Elite: Ironies and Unfulfilled Promises*. Rowman & Littlefield, 2018; Zweigenhaft, Richard L. "The 116th Congress has more women and people of color than ever – but there's still room to improve." The Conversation. November 8, 2018. https://theconversation.com/the-116th-congress-has-more-women-and-people-of-color-than-ever-but-theres-still-room-to-improve-105930; Hanks, Angela, Danyelle Solomon, and Christian E. Weller. "Systematic inequality: How America's structural racism helped create the black-white wealth gap." *Washington: Center for American Progress* (2018). https://www.americanprogress.org/issues/race/reports/2018/02/21/447051/systematic-inequality/.
[36] Osseo-Asare, Abena Dove. *Bitter roots: The search for healing plants in Africa*. University of Chicago Press, 2014. See also: Tilley, Helen. *Africa as a living laboratory empire, development, and the problem of scientific knowledge, 1870-1950*. Chicago, Ill: University of Chicago Press, 2013; Fullwiley, Duana. *The enculturated gene: sickle cell health politics and biological difference in West Africa*. Princeton: Princeton University Press, 2011.
[37] Schweiker, William, ed. *The Blackwell companion to religious ethics*. John Wiley & Sons, 2008.
[38] Zigon, Jarrett. *Morality: an anthropological perspective*. Berg, 2008; Lambek, Michael, ed. *Ordinary Ethics: Anthropology, language, and action*. Fordham University Press, 2010; Fassin, Didier and Samuel Lézé, eds. *Moral anthropology. A critical reader.* London: Routledge, 2014; Durkheim, Emile. "The determination of moral facts." *Sociology and philosophy* (1953): 35-62.
[39] Ibid.
[40] Bourdieu, Pierre. *The logic of practice*. Stanford University Press, 1990; Bourdieu, Pierre, and Loïc Wacquant. *An invitation to reflexive sociology*. University of Chicago Press, 1992.
[41] Dutton, Tim. "An Overview of National AI Strategies." *Medium*, July 13, 2018. https://medium.com/politics-ai/an-overview-of-national-ai-strategies-2a70ec6edfd
[42] Hern, Alex. "Facebook translates 'good morning' into 'attack them', leading to arrest." *The Guardian*, October 24, 2017. https://www.theguardian.com/technology/2017/oct/24/facebook-palestine-israel-translates-good-morning-attack-them-arrest
[43] Ochigame, Rodrigo, and James Holston. "Filtering Dissent: Social Media and Land Struggles in Brazil." New Left Review. (2016) 99: 85-110.
[44] Ess, Charles. "Lost in translation?: Intercultural dialogues on privacy and information ethics." Ethics and Information Technology (2005)1: 1-6; Lü Yao-Huai "Privacy and data privacy issues in contemporary China." Ethics and Information Technology (2005)1: 7-15; McDougall, Bonnie and Hansson, Anders. *Chinese concepts of privacy.* Boston: Brill, 2002; Wu, Yanfang, Tuenyu Lau, David J. Atkin, and Carolyn A. Lin. "A comparative study of online privacy regulations in the US and China." *Telecommunications Policy* 35, no. 7 (2011): 603-616.
[45] Schauer, Frederick. "The social construction of privacy." In LaMay, Craig, ed. *Journalism and the debate over privacy*, pp. 15-28. Routledge, 2003.
[46] Appadurai, Arjun. "Disjuncture and difference in the global cultural economy." *Theory, culture & society* 7, no. 2-3 (1990): 295-310; Tsing, Anna Lowenhaupt. *Friction: An ethnography of global connection*. Princeton University Press, 2011.
[47] For a discussion of traveling technologies in a human rights context, see: Hagerty, Alexa. "A Lovely Grave for Learning": Exhuming the Dead of Guatemala's Genocide. Los Angeles Review of Books, June 2018.
[48] For a discussion of "frictions and flows" in the context of resource frontiers in Asia, see: Rubinov, Igor. "Growing at the Margins: Enlivening a Neglected Post-Soviet Frontier." *Frontier Assemblages: The Emergent Politics of Resource Frontiers in Asia* (2019): 99-115.
[49] Madhavan, Samir. "How an Artificial Intelligence project is helping to make Brazilian politicians more accountable" HackerNoon, April 2, 2018. https://hackernoon.com/how-an-artificial-intelligenceproject-is-helping-to-make-brazilian-politicians-more-accountable-d0178895988c

[50] Basu, Arindrajit and Abhijit Majumdar. "Reimagining the possible in the Indian healthcare ecosystem with emerging technologies" PwC, 2018. https://www.pwc.in/assets/pdfs/publications/2018/reimagining-the-possible-in-the-indian-healthcare-ecosystem-with-emerging-technologies.pdf
[51] Zenvus. 2018. "Zenvus – Intelligent Solutions for Farms and Gardens." https://www.zenvus.com/.
[52] Basu, Arindrajit and Elonnai Hickok. "Artificial Intelligence in the Governance Sector in India." The Centre for Internet and Society, India. 2018 https://cis-india.org/internet- governance/ai-and-governance-case-study-pdf
[53] In addition to counting the number of unique views, the software also infers gender, age, and emotional state: "happy, unsatisfied, surprised, or neutral." Amigo, Ignacio. "The Metro Stations of São Paulo That Read Your Face." CityLab, May 8, 2018 https://www.citylab.com/design/2018/05/the-metro-stations-of-sao-paulo-that-read-your-face/559811/; See also Via Quatro website: http://www.viaquatro.com.br/a-via-quatro
[54] Waldron, Travis. "How Police Violence Paves the Way to Authoritarianism." Huffington Post. February 23, 2019. https://www.huffpost.com/entry/brazil-police-violence-jair-bolsonaro_n_5c475d0ae4b027c3bbc61a54
[55] The Indian government currently has plans to bring predictive policing technologies in 5 states: Kerala, Odisha, Maharashtra, Haryana, and Tripura. Basu, Arindrajit and Elonnai Hickok. "Artificial Intelligence in the Governance Sector in India." The Centre for Internet and Society, India. 2018 https://cis-india.org/internet- governance/ai-and-governance-case-study-pdf.
[56] Unique Identification Authority of India, Aadhaar dashboard. Accessed January 10, 2019. uidai.gov.in/aadhaar_dashboard/index.php.
[57] Gelb, Alan and Anna Diofasi. "Biometric Elections in Poor Countries: Wasteful or a Worthwhile Investment?" Working Paper 435, Center for Global Development, 2016. https://www.cgdev.org/publication/biometric-elections-poor-countries-wasteful-or-worthwhile-investment
[58] Naman M. Aggarwal, Wafa Ben-Hassine, and Raman Jit Singh Chima. "National digital identity programmes: what's next?" Access Now. March 2018. https://www.accessnow.org/cms/assets/uploads/2018/06/Digital-Identity-Paper-2018-05.pdf
[59] The Electronic Frontier Foundation describes such centralized databases as a "honeypot of sensitive data vulnerable to exploitation." https://www.eff.org/issues/national-ids. See also: Panday, Jyoti. "Can India's Biometric Identity Program Aadhaar Be Fixed?" Electronic Frontier Foundation. February 27, 2018. https://www.eff.org/deeplinks/2018/02/can-indias-aadhaar-biometric-identity-program-be-fixed
[60] "The government claims Aadhaar is not mandatory to access healthcare. But HIV patients say they are being coerced into submitting their numbers." Rao, Menaka. "Why Aadhaar is prompting HIV positive people to drop out of treatment programmes across India." Scroll.In, November 17, 2017. https://scroll.in/pulse/857656/across-india-hiv-positive-people-drop-out-of-treatment-programmes-as-centres-insist-on-aadhaar
[61] The Wire. "Of 42 'Hunger-Related' Deaths Since 2017, 25 'Linked to Aadhaar Issues'" September 21, 2018. thewire.in/rights/of-42-hunger-related-deaths-since-2017-25-linked-to-aadhaar-issues.
[62] Sullivan, Clare. "Digital citizenship and the right to digital identity under international law." *Computer Law & Security Review* 32 (2016): 474-481.
[63] Barocas, Solon, and Andrew D. Selbst. "Big data's disparate impact." *California Law Review* 104 (2016): 671; Williams, Betsy Anne, Catherine F. Brooks, and Yotam Shmargad. "How algorithms discriminate based on data they lack: Challenges, solutions, and policy implications." *Journal of Information Policy* 8 (2018): 78-115; Garvie, Clare. *The perpetual line-up: Unregulated police face recognition in America*. Georgetown Law, Center on Privacy & Technology, 2016. https://www.perpetuallineup.org/
[64] Dastin, Jeffrey. "Amazon scraps secret AI recruiting tool that showed bias against women." *Reuters.* October 9 2018.
[65] Bartlett, Robert, Adair Morse, Richard Stanton, and Nancy Wallace. "Consumer-Lending Discrimination in the Era of FinTech." *Unpublished working paper. University of California, Berkeley* (2018); Glantz, Aaron and Emmanuel Martinez. "Kept out: How banks block people of color from homeownership." *Associated Press.* February 15, 2018. https://apnews.com/ae4b40a720b74ad8a9b0bfe65f7a9c29; Waddell, Kaveh. "How Algorithms Can Bring Down Minorities' Credit Scores." The Atlantic. December 2, 2016. https://www.theatlantic.com/technology/archive/2016/12/how-algorithms-can-bring-down-minorities-credit-scores/509333/
[66] Caliskan, Aylin, Joanna J. Bryson, and Arvind Narayanan. "Semantics derived automatically from language corpora contain human-like biases." *Science* 356, no. 6334 (2017): 183-186; Zhao, Jieyu, Tianlu Wang, Mark Yatskar, Vicente Ordonez, and Kai-Wei Chang. "Men also like shopping: Reducing gender bias amplification using corpus-level constraints." *arXiv preprint arXiv:1707.09457* (2017); Bolukbasi, Tolga, Kai-Wei Chang, James Y.

Zou, Venkatesh Saligrama, and Adam T. Kalai. "Man is to computer programmer as woman is to homemaker? debiasing word embeddings." In *Advances in Neural Information Processing Systems*, pp. 4349-4357. 2016; Thompson, Andrew. "Google's Sentiment Analyzer Thinks Being Gay Is Bad." *Motherboard* October 25, 2017. https://motherboard.vice.com/en_us/article/j5jmj8/google-artificial-intelligence-bias.

[67] World Economic Forum. "How to Prevent Discriminatory Outcomes in Machine Learning." March, 2018. http://www3.weforum.org/docs/WEF_40065_White_Paper_How_to_Prevent_Discriminatory_Outcomes_in_Machine_Learning.pdf.

[68] The World Bank Poverty and Shared Prosperity 2018 http://www.worldbank.org/en/publication/poverty-and-shared-prosperity; Transparency International Corruption Perceptions Index 2018 https://www.transparency.org/cpi2018; The Economist Intelligence Unit Democracy Index 2018 https://www.eiu.com/topic/democracy-index.

[69] Nebehay, Stephanie. "U.N. calls for Myanmar generals to be tried for genocide, blames Facebook for incitement." *Reuters.* August 27, 2018. https://www.reuters.com/article/us-myanmar-rohingya-un/myanmar-generals-had-genocidal-intent-against-rohingya-must-face-justice-u-n-idUSKCN1LC0KN
Nebehay 2018; Stecklow 2018; Stecklow, Steve. "Why Facebook is losing the war on hate speech in Myanmar." *Reuters.* August 15, 2018. https://www.reuters.com/investigates/special-report/myanmar-facebook-hate/

[70] Brown, Adrian. "In China's Xinjiang, surveillance is all pervasive." Al Jazeera Blog, 6 Nov 2018. https://www.aljazeera.com/blogs/asia/2018/11/china-xinjiang-surveillance-pervasive-181106010716601.html; Samuel, Sigal. "China Is Going to Outrageous Lengths to Surveil Its Own Citizens." The Atlantic. August 16, 2018. https://www.theatlantic.com/international/archive/2018/08/china-surveillance-technology-muslims/567443/; Freedom House. "The Battle for China's Spirit Religious Revival, Repression, and Resistance under Xi Jinping." A Freedom House Special Report, November 2017. https://freedomhouse.org/sites/default/files/FH_ChinasSprit2017_Abridged_FINAL_compressed.pdf; Human Rights Watch. "World Report 2018." January 9, 2018. https://www.hrw.org/world-report/2018#

[71] Aziza, Sarah. "Jamal Khashoggi Wasn't the First—Saudi Arabia Has Been Going After Dissidents for Decades." *The Intercept.* October 12 2018; Kirkpatrick, David D. "Israeli Software Helped Saudis Spy on Khashoggi, Lawsuit Says." *The New York Times.* December 2, 2018.

[72] Garvie, Clare. *The perpetual line-up: Unregulated police face recognition in America*. Georgetown Law, Center on Privacy & Technology, 2016. https://www.perpetuallineup.org; Angwin, Julia, Jeff Larson, Surya Mattu, and Lauren Kirchner. "Machine bias: There's software used across the country to predict future criminals." *And it's biased against blacks. ProPublica* (2016). https://www.propublica.org/article/machine-bias-risk-assessments-in-criminal-sentencing.

[73] Burt, Chris. "Possibility of Chinese facial biometrics systems in Brazilian CCTV network raises concerns." Biometric Update. January 21, 2019. https://www.biometricupdate.com/201901/possibility-of-chinese-facial-biometrics-systems-in-brazilian-cctv-network-raises-concerns; Walsh, Ray. "Brazil employs facial recognition technology to tackle crime hotpots." https://www.bestvpn.com/privacy-news/brazil-facial-recognition-cameras/

[74] Ibid.

[75] Poushter, Jacob, Caldwell Bishop and Hanyu Chwe. "Social Media Use Continues to Rise in Developing Countries but Plateaus Across Developed Ones" A Pew Center Report, June 19, 2018. http://www.pewglobal.org/2018/06/19/social-media-use-continues-to-rise-in-developing-countries-but-plateaus-across-developed-ones/

[76] Bughin, Jacques, Jeongmin Seong, James Manyika, Michael Chui, and Raoul Joshi. "Notes from the AI frontier: Modeling the impact of AI on the world economy." Discussion Paper, McKinsey Global Institute, September 2018. https://www.mckinsey.com/featured-insights/artificial-intelligence/notes-from-the-AI-frontier-modeling-the-impact-of-ai-on-the-world-economy

[77] Bandwidth inequality, closely linked to income inequality, is key factor. As of 2014, 3 countries hosted 50% of the globally installed bandwidth potential, and 10 countries almost 75%. Hilbert, Martin. "The bad news is that the digital access divide is here to stay: domestically installed bandwidths among 172 countries for 1986–2014." *Telecommunications Policy* 40, no. 6 (2016): 567-581; Only 53.6% of all households have internet access. In emerging countries, the share is only 15%. International Telecommunication Union, "ICT Development Index 2017" http://www.itu.int/net4/ITU-D/idi/2017/; WIPO Technology Trends – Artificial Intelligence 2019.

[78] World Economic Forum, "Global Gender Gap Report 2018." http://reports.weforum.org/global-gender-gap-report-2018/assessing-gender-gaps-in-artificial-intelligence/?doing_wp_cron=1551634743.8178520202636718750000EY Global 2018.

[79] World Economic Forum, “Networked Readiness Index” in The Global Information Technology Report 2016 http://reports.weforum.org/global-information-technology-report-2016/networked-readiness-index/
[80] PwC Global Artificial Intelligence Study 2017. “Sizing the Prize.” https://www.pwc.com/gx/en/issues/data-and-analytics/publications/artificial-intelligence-study.html
[81] Nedelkoska, L. and G. Quintini (2018), “Automation, skills use and training.” *OECD Social, Employment and Migration Working Papers*, No. 202, OECD Publishing, Paris, https://doi.org/10.1787/2e2f4eea-en; World Economic Forum, “Global Information Technology Report 2016.” http://reports.weforum.org/global-information-technology-report-2016/networked-readiness-index/; International Labour Organization. “Work for a brighter future.” ILO Global Commission on the Future of Work Report, January 2019. http://www.ilo.org/global/publications/books/WCMS_662410/lang--en/index.htm
[82] International Labour Organization. “Work for a brighter future.” ILO Global Commission on the Future of Work Report, January 2019. http://www.ilo.org/global/publications/books/WCMS_662410/lang--en/index.htm
[83] PwC Global Artificial Intelligence Study 2017. “Sizing the Prize: Exploiting the AI Revolution.” https://www.pwc.com/gx/en/issues/data-and-analytics/publications/artificial-intelligence-study.html; United Nations Conference on Trade and Development (UNCTAD). Policy Brief No. 50. “Robots and industrialization in developing countries.” 2016. https://unctad.org/en/PublicationsLibrary/presspb2016d6_en.pdf; Lee, Kai-Fu. “AI Could Devastate the Developing World.” Bloomberg. September 17, 2018. https://www.bloomberg.com/opinion/articles/2018-09-17/artificial-intelligence-threatens-jobs-in-developing-world
[84] Bruckner, Mattias, Marcelo LaFleur, and Ingo Pitterle. “The impact of the technological revolution on labour markets and income distribution.” July 31, 2017. United Nations, Department of Economic & Social Affairs. https://www.un.org/development/desa/dpad/wpcontent/uploads/sites/45/publication/2017_Aug_Frontier-Issues-1.pdf
[85] Nedelkoska, Ljubica and Quintini, Glenda (2018), "Automation, skills use and training.” *OECD Social, Employment and Migration Working Papers*, No. 202, OECD Publishing, Paris, https://doi.org/10.1787/2e2f4eea-en; World Economic Forum, “Global Information Technology Report 2016.” http://reports.weforum.org/global-information-technology-report-2016/networked-readiness-index/
[86] International Labour Organization. “Work for a brighter future.” ILO Global Commission on the Future of Work Report, January 2019. http://www.ilo.org/global/publications/books/WCMS_662410/lang--en/index.htm
[87] Lee, David. “Why Big Tech pays poor Kenyans to teach self-driving cars.” BBC. November 3, 2018. https://www.bbc.com/news/technology-46055595. For an excellent discussion of the labor and environmental impacts of the Amazon Alexa system, see Crawford, Kate and Joler, Vladan. “Anatomy of an AI system” 2018. https://anatomyof.ai/
[88] Internet Society, “Paths to Our Digital Future.” Global Internet Report 2017. https://future.internetsociety.org/2017/wp-content/uploads/2017/09/2017-Internet-Society-Global-Internet-Report-Paths-to-Our-Digital-Future.pdf
[89] Stiglitz, Joseph E. *The price of inequality: How today's divided society endangers our future*. WW Norton & Company, 2012; Piketty, Thomas. *Le capital au XXIe siècle*. Le Seuil, 2013; World Economic Forum, “Global Risks Report 2019” https://www.weforum.org/reports/the-global-risks-report-2019; Oxfam Report, “Working for the Few: Political capture and economic inequality.” January 2014. https://www.oxfam.org/sites/www.oxfam.org/files/bp-working-for-few-political-capture-economic-inequality-200114-summ-en.pdf; Human Rights Watch. “World Report 2018.” January 9, 2018. https://www.hrw.org/world-report/2018#
[90] Ibid.
[91] United Nations News, “UN chief urges top digital tech panel to come up with ‘bold, innovative ideas’ for an 'inclusive' future.” January 23, 2019. https://news.un.org/en/story/2019/01/1031092
[92] The World Economic Forum, “Global Risks Report 2019” has flagged AI-driven surveillance as a troubling form of social control and a threat to democracy: “Global politics will be affected: authoritarianism is easier in a world of total visibility and traceability, while democracy may turn out to be more difficult—many societies are already struggling to balance threats to privacy, trust and autonomy against promises of increased security, efficiency and novelty. Geopolitically, the future may hinge in part on how societies with different values treat new reservoirs of data.” https://www.weforum.org/reports/the-global-risks-report-2019
[93] Sterman, David. “How the Government Exaggerated the Successes of NSA Surveillance.” Slate. January 16, 2014. http://www.slate.com/blogs/future_tense/2014/01/16/nsa_surveillance_how_the_government_exaggerated_the_way_its_programs_stopped.html
[94] https://www.wsj.com/articles/with-pop-star-as-bait-china-nabs-suspects-using-facial-recognition-1527004150

[95] Wong, Chun Han. "With Pop Star as Bait, China Nabs Suspects Using Facial Recognition." Wall Street Journal. May 22, 2018 https://www.nytimes.com/2018/07/08/business/china-surveillance-technology.html
[96] Ibid.
[97] Freedom House rates Singapore as having a net status of a "partly free" society with a "not free" press. https://freedomhouse.org/report/freedom-press/2017/singapore; Privacy International describes Singapore as an "endemic surveillance society." https://www.privacyinternational.org/article.shtml?cmd[347]=x-347-559597; The Economist Intelligence Unit Democracy Index 2018, rates Singapore as a "hybrid regime." https://www.eiu.com/topic/democracy-index.
[98] Jiow, Hee Jhee, and Sofia Morales. "Lateral Surveillance in Singapore." *Surveillance & Society* 13, no. 3/4 (2015): 327.
[99] Mandavilli, Apoorva. "SARS epidemic unmasks age-old quarantine conundrum." Nature Medicine volume 9, page 487 (2003); Tan, Chorh-Chuan. "SARS in Singapore-key lessons from an epidemic." *Annals-Academy of Medicine Singapore* 35, no. 5 (2006): 345; World Health Organization. "Severe acute respiratory syndrome – Singapore, 2003." *Weekly Epidemiological Record*, *78*(19), 159. http://www.who.int/docstore/wer/pdf/2003/wer7819.pdf; Chua, M. H. (2004). *A defining moment: How Singapore beat SARS*. Singapore: Institute of Policy Studies, pp. 27, 67, 194–197. World Health Organization. (2004); Power, John and Xiaosui Xiao, eds. *The Social Construction of SARS: Studies of a health communication crisis.* John Benjamins Publishing, 2008.
[100] Ibid.
[101] Harris, Shane. "The Social Laboratory." *Foreign Policy*. July 29, 2014. https://foreignpolicy.com/2014/07/29/the-social-laboratory/
[102] Berwick, Angus. "How ZTE helps Venezuela create China-style social control." Reuters. November 14, 2018. www.reuters.com/investigates/special-report/venezuela-zte/.
[103] Ibid.
[104] Mann, Daniel. 2019. "'I Am Spartacus': individualising visual media and warfare." *Media, Culture & Society.* 41 (1): 38-53. See also, "Jerusalem: A Palestinian woman and a member of the Israeli security forces photograph each other." Photograph: Ahmad Gharabli/AFP/Getty Images. https://www.theguardian.com/artanddesign/gallery/2017/dec/23/the-20-photographs-of-the-week
[105] Harding, Luke. "Saudi Arabia criticised over text alerts tracking women's movements." The Guardian, November 23, 2012. https://www.theguardian.com/world/2012/nov/23/saudi-arabia-text-alerts-women
[106] Harris, Shane. 2014. "The Social Laboratory." *Foreign Policy*. July 29 https://foreignpolicy.com/2014/07/29/the-social-laboratory/
[107] Knight, Will. "Why does Beijing suddenly care about AI ethics?" *MIT Technology Review*. May 31, 2019. https://www.technologyreview.com/s/613610/why-does-china-suddenly-care-about-ai-ethics-and-privacy/
[108] State Council Notice Concerning Issuance of the Planning Outline for the Establishment of a Social Credit System (2014-2020).
[109] Ohlberg, Mareike, Ahmed, Shazeda and Lang, Bertram. 2017. Central Planning, Local Experiments: The Complex Implementation of China's Social Credit System. Mercator Institute for China Studies. https://www.merics.org/sites/default/files/2018-03/171212_China_Monitor_43_Social_Credit_System_Implementation_1.pdf
[110] Kostka, Genia. 2018. China's Social Credit Systems and Public Opinion: Explaining High Levels of Approval. https://papers.ssrn.com/sol3/papers.cfm?abstract_id=3215138.
[111] Crawford, Kate and Ryan Calo. "There is a blind spot in AI research." *Nature* (2016): 538:311–313
[112] https://datasociety.net/research/ai-on-the-ground/
[113] Eubanks, Virginia. "Want to Predict the Future of Surveillance? Ask Poor Communities." American Prospect Magazine, January 15, 2014. https://prospect.org/article/want-predict-future-surveillance-ask-poor-communities. See also: Eubanks, Virginia. *Automating inequality: How high-tech tools profile, police, and punish the poor*. St. Martin's Press, 2018.
[114] Roth, John K. *The Failures of Ethics: Confronting the Holocaust, Genocide, and Other Mass Atrocities*. Oxford University Press, USA, 2015.
[115] Floridi, Luciano. "Soft Ethics and the Governance of the Digital." *Philosophy & Technology* 31, no. 1 (2018):1-8.
[116] Stewart, Uyi. "AI can help to bridge the digital divide and create an inclusive society." ITU News. April 23, 2018. https://news.itu.int/ai-can-help-to-bridge-the-digital-divide-and-create-an-inclusive-society/
[117] ibid.